\documentclass[10pt,conference]{IEEEtran}
\usepackage{comment}
\usepackage{tikz}
\usepackage{tikzducks}
\usepackage{ducksay}
\usepackage{xspace} 

\usepackage{caption}
\usepackage{multirow}
\usepackage{graphics}
\usepackage{comment}
\usepackage{url}
\usepackage{framed}
\usepackage{ifthen}

\usepackage{hyperref}
\usepackage{color}

\newboolean{showcomments}
\setboolean{showcomments}{false}
\ifthenelse{\boolean{showcomments}}
  {\newcommand{\nb}[2]{
  \fbox{\bfseries\sffamily\scriptsize#1}
    {\sf\small$\blacktrianglerig$\textit{\textcolor{green}{#2}}$\blacktriangleleft$}
   }
  }
  {\newcommand{\nb}[2]{}
   
  }
\definecolor{patriarch}{rgb}{0.5, 0.0, 0.5}
\definecolor{oceanboatblue}{rgb}{0.0, 0.47, 0.75}

\usepackage[framemethod=tikz]{mdframed}
\newcounter{finding}
\newcommand{\finding}[1]{\refstepcounter{finding}
  \vspace{1mm}
 \begin{mdframed}[linecolor=gray,roundcorner=12pt,backgroundcolor=gray!15,linewidth=3pt,innerleftmargin=2pt, leftmargin=0cm,rightmargin=0cm,topline=false,bottomline=false,rightline = false]
  \textbf{Finding \arabic{finding}:} #1
 \end{mdframed}
 \vspace{-0.3cm}
}

\newcommand{\originalp}{Chen et al. \cite{chen2016software}\xspace}
\newcommand{\originalcite}{\cite{chen2016software}\xspace}

\ifCLASSOPTIONcompsoc
  \usepackage[nocompress]{cite}
\else
  \usepackage{cite}
\fi

\ifCLASSINFOpdf
\else
\fi

\begin{document}

\title{On The Effectiveness of One-Class Support Vector Machine in Different Defect Prediction Scenarios}


\author{\IEEEauthorblockN{Rebecca Moussa}
\IEEEauthorblockA{\textit{Department of Computer Science} \\
\textit{University College London (UCL)}\\
London, United Kingdom \\
r.moussa@ucl.ac.uk}
\and
\IEEEauthorblockN{Danielle Azar}
\IEEEauthorblockA{\textit{Computer Science and Mathematics Department} \\
\textit{Lebanese American University}\\
Byblos, Lebanon\\
danielle.azar@lau.edu.lb}
\and
\IEEEauthorblockN{Federica Sarro}
\IEEEauthorblockA{\textit{Department of Computer Science} \\
\textit{University College London (UCL)}\\
London, United Kingdom \\
f.sarro@ucl.ac.uk}
}

\IEEEtitleabstractindextext{
\begin{abstract}
Defect prediction aims at identifying software components that are likely to cause faults before a software is made available to the end-user. To date, this task has been modeled as a two-class classification problem, however its nature also allows it to be formulated as a one-class classification task. 
Previous studies show that One-Class Support Vector Machine (OCSVM) can outperform two-class classifiers for within-project defect prediction, however it is not effective when employed at a finer granularity (i.e., commit-level defect prediction). In this paper, we further investigate whether learning from one class only is sufficient to produce effective defect prediction model in two other different scenarios (i.e., granularity), namely cross-version and cross-project defect prediction models, as well as replicate the previous work at within-project granularity for completeness. 
Our empirical results confirm that OCSVM performance remain low at different granularity levels, that is, it is outperformed by the two-class Random Forest (RF) classifier for both cross-version and cross-project defect prediction.
While, we cannot conclude that OCSVM is the best classifier, our results still show interesting findings. While OCSVM does not outperform RF, it still achieves performance superior to its two-class counterpart (i.e., SVM) as well as other two-class classifiers studied herein. We also observe that OCSVM is more suitable for both cross-version and cross-project defect prediction, rather than for within-project defect prediction, thus suggesting it performs better with heterogeneous data. 
We encourage further research on one-class classifiers for defect prediction as these techniques may serve as an alternative when data about defective modules is scarce or not available.
\end{abstract}

}

\maketitle

\IEEEdisplaynontitleabstractindextext

%
\IEEEpeerreviewmaketitle

\section{Introduction}
\label{sec:introduction}

\IEEEPARstart{S}oftware defects can be very costly to both, the users and the company providing the software. 
As the world becomes more dependent on software, detecting defects early in the development process becomes more and more critical: the earlier a defect is found and fixed, the less it costs\cite{Capers2008}.

The research in software defect prediction aims at the automatic and early identification of problematic software components, in order to direct the most of the testing effort towards them.

A large number of classifiers have been investigated to build software defect prediction models\cite{hall2011systematic}. The most commonly used type of these classification models, two-class predictors, rely on the use of training data consisting of instances of the two classes being studied (i.e., defective and non-defective instances). These include widely known models such as Random Forest (RF) \cite{breiman2001random}, Support Vector Machines (SVM) \cite{burges1998tutorial}, Naïve Bayes (NB) \cite{witten2005practical}, etc. The use of both classes allows such models to learn characteristics about both types of modules possibly making the task of predicting a new unseen module more accurate. 

One-class predictors are another type of classification models which have been recently gaining more attention. 
These techniques require the availability of one class only (i.e., non-defective instances) in order to learn characteristics of the training data. An unseen module is classified as an outlier if its characteristics are very different from those learnt by the model from the training set and hence it does not lie within the boundaries created by the technique. 

The fact that the number of defective modules in real world systems is much lower than the non-defective ones, leads to  highly imbalanced datasets that often cause two-class predictors to produce poor results \cite{hall2011systematic,yao2020assessing}. 
As a result, studies have attempted to improve the performance of defective prediction models by applying  well-known under- and over-sampling approaches like Random Under Sampling (RUS), Synthetic Minority Over-sampling TEchnique (SMOTE) \cite{chawla2002smote}, ADAptive SYNthetic sampling technique (ADASYN) \cite{he2008adasyn}, and SMOTUNED \cite{agrawal2018better} to balance the datasets. However, in a recent study investigating the impact of class rebalancing techniques on the performance measures and interpretation of defect models, Tantithamthavorn et al. \cite{tantithamthavorn2018impact} show that sampling affects the interpretability of defect prediction models and it should be avoided when deriving knowledge and insights from them. 
Thus, the availability of a model that only requires non-defective instances and no data balancing to learn and accurately classify defective modules could offer an alternative to address data-imbalance. 

While several studies have shown that one-class predictors can be successfully used to address various classification tasks suffering from imbalanced data \cite{manevitz2001one, erfani2016high, zhang2006fall, li2010positive, khan2014one}, only two investigations have been carried out on the use of one-class SVM (OCSVM) for software defect prediction showing promising results when using OCSVM for within-project defect prediction \originalcite by the NASA datasets, and negative results when it is used for just-in-time defect prediction \cite{10011488}.

In this paper, we further investigate the use of OCSVM in different prediction scenarios (namely, within-project, cross-version and cross-project) in the light of the more recent advances in defect prediction empirical research including the use of more recent and robust data, hyper-parameter tuning, robust evaluation measures and statistical tests.
Specifically, we investigate both the canonical OCSVM, as proposed in the literature, as well as a novel way to perform OCSVM hyper-parameter tuning by using a minimal number of defective instances (we refer to this as OCSVM$_T$).

Our results show that, while OCSVM performs generally better than naive baseline, its two-class counterpart SVM, and some other two-class defect predictors for some scenarios, it is not the best approach for all scenarios.  In fact, it is outperformed by Random Forest. While we cannot recommend the use of OCSVM for the within-project scenario, therefore confuting the results of previous study \originalcite, its performance notably improves when heterogeneous data is used, as for the cross-project scenario, where OCSVM$_T$ is able to always achieve statistically significantly better estimates with respect to three widely used two-class approaches (i.e., Support Vector Machine, Naive Bayes, Logistic Regression), while it is able to outperform Random Forest with statistical significance in 22\% of the cases for the within-project and cross-version scenarios, and 44\% of the cases for the cross-project scenario.

As these results confound the findings by \originalp, for due diligence, we also conduct a replication of their work, in which we follow as closely as possible the procedure undertaken by the authors \cite{shull2008role}. To this end, we use their same datasets and validation procedure (as summarised in Table \ref{tab:studies2}), and at the same time enrich the experimental design by using a robust evaluation measure, statistical significance test, and three benchmarks. Our results still do not confirm the conclusion of the prior work \originalcite \space as we found that while OCSVM generally performs similarly to its two-class counterpart, and it does not outperform the best two-class classifier (i.e., Random Forest).

Although we cannot conclude that OCSVM is suitable for all scenarios, we believe reporting negative  findings still helps advance the research agenda in defect prediction. Properly conducted studies with negative, null or neutral results are essential for the progression of science and its self-correcting nature as much as positive results are \cite{paige2017foreword, tichy2000hints, ferguson2012vast}. Sharing these findings prevents other researchers from following the same route and it can help them adjust their own research plans, thus saving time and effort; it can also provide researchers with the knowledge needed to develop alternative strategies and evolve new better ideas \cite{paige2017foreword, menzies2017negative}.

In fact, even though our study shows that OCSVM does not consistently outperform all two-class classifiers considered herein, it also provides some interesting initial evidence on the potential advantages of using a minimal number of defective instances for hyper-parameter tuning, especially in case of heterogeneous data (i.e., training data composed by different versions of the same target project, or from projects different from the target one). 
Thus, we encourage future work to further explore the extent to which using different ratio of defective instances coming from the target project or different ones, can impact the prediction performance, as this can serve as alternative solutions when data on defective instances is scarce or not available.

\begin{table}[tb]
\centering
\resizebox{0.85\linewidth}{!}{
\begin{tabular}{llr}
Repository                       & Dataset               & No. of modules (faulty \%) \\ \hline
\multirow{6}{*}{NASA \cite{petric2016jinx}}       & CM1                   & 296 (12.84\%)              \\
                            & KC3                   & 123 (13.00\%)              \\
                            & MW1                   & 253 (10.67\%)              \\
                            & PC1                   & 661 (7.87\%)               \\
                            & PC3                   & 1043 (12.18\%)             \\
                            & PC5                   & 94 (19.15\%)               \\ \hline
\multirow{18}{*}{Realistic\cite{yatish2019mining}} & Activemq\_5.3.0       & 2367 (10.90\%)             \\
                            & Activemq\_5.8.0       & 3420 (6.02\%)              \\
                            & Camel\_2.10.0         & 7914 (2.91\%)              \\
                            & Camel\_2.11.0         & 8846 (2.17\%)              \\
                            & Derby\_10.3.1.4       & 2206 (30.33\%)             \\
                            & Derby\_10.5.1.1       & 2705 (14.16\%)             \\
                            & Groovy\_1\_6\_BETA\_1 & 821 (8.53\%)               \\
                            & Groovy\_1\_6\_BETA\_2 & 884 (8.60\%)               \\
                            & Hbase\_0.95.0         & 1669 (22.95\%)             \\
                            & Hbase\_0.95.2         & 1834 (26.34\%)             \\
                            & Hive\_0.10.0          & 1560 (11.28\%)             \\
                            & Hive\_0.12.0          & 2662 (8.00 \%)             \\
                            & Jruby\_1.5.0          & 1131 (7.25\%)              \\
                            & Jruby\_1.7.0          & 1614 (5.39\%)              \\
                            & Lucene\_3.0           & 1337 (11.59\%)             \\
                            & Lucene\_3.1           & 2806 (3.81\%)              \\
                            & Wicket\_1.3.0-beta2   & 1763 (7.37\%)              \\
                            & Wicket\_1.5.3         & 2578 (4.07\%)   \\ \hline 
\end{tabular}}
\caption{\textbf{Datasets}. Total number of modules and percentage of faulty components per dataset.}
\label{tab:stat-rep-datasets}
\end{table}

\label{sec:ConcepRep}

\section{Empirical Study Design}
\label{sec:concep-design}

\subsection{Research Questions}
\label{sec:concep-design-rqs}

First and foremost, we investigate if OCSVM is able to outperform a naive baseline. 
Recent studies have stressed on the importance of including a baseline benchmark to assess any newly proposed prediction models \cite{d2012evaluating, sarro2018linear}. This check is essential to assess whether OCSVM is able to learn and differentiate non-defective modules from defective ones, instead of randomly classifying them. For this reason, we pose our first research question:

\textbf{RQ1. OCSVM vs. Random}: Does OCSVM outperform a Random classifier?

\noindent In order to answer RQ1, we compare OCSVM with a basic Random Classifier, which is completely independent of the training data (i.e., there is no learning phase), and instead generates predictions uniformly at random \cite{scikitLearn}. Any prediction system must outperform the Random Classifier, otherwise this would indicate that the prediction system is not learning any information from the training data \cite{sarro2018linear}.

Our second benchmark consists in assessing whether OCSVM performs better than its two-class counterpart, SVM. This is a required check, as if the results reveal the opposite, there is no advantage of using the one-class classifier. To this end, we ask: 

\textbf{RQ2. OCSVM vs. SVM}: Does OCSVM outperform SVM, its two-class counterpart? 

To answer this question we compare OCSVM versus its two-class version by using the same kernel.
We use SVM both out-of-the-box and tuned, as the former has been widely used in past studies, although we discourage its use as lack of proper hyper-parameter tuning might lead to less accurate predictions \cite{tantithamthavorn2018impact, sarro2012further, di2011genetic}.

A positive answer to RQ2 means that using only information about the non-defective class is sufficient to achieve accurate predictions for SVM. However, it might still happen that OCSVM is not comparable with other two-class classifiers. This leads us to our third, and last research question where we compare the performance of OCSVM to that of other well-known traditional two-class techniques: 

\textbf{RQ3. OCSVM vs. Traditional ML}: Does OCSVM outperform traditional machine learning techniques?

\noindent To address this question, we compare OCSVM with three traditional two-class machine learning techniques, namely NB, LR and RF, which have been widely used in defect prediction studies \cite{hall2011systematic}. Similarly to RQ2, we experiment with both tuned and non-tuned versions.

In the remaining of this Section we describe in details the experimental setting used to answer these RQs.


\subsection{Datasets}
\label{sec:data}
In our empirical study we have used two sets of publicly available software project datasets: NASA \cite{petric2016jinx} and the Realistic\cite{yatish2019mining} datasets. 


\subsubsection{NASA dataset}
The NASA datasets, made publicly available by the NASA’s
Metrics Data Program (MDP), contain data on the NASA Lunar space system software written both in C and Java.
In our empirical study we use the NASA datasets curated by Petrić et al. \cite{petric2016jinx}, who applied rules to clean and remove the erroneous data contained in the original NASA datasets \cite{petric2016jinx, shepperd2013data}. 
We use the six datasets listed in Table \ref{tab:stat-rep-datasets}. 
They contain static code measures (e.g., LOC, Halstead, MaCabe) and the number of defects for each software component. 


\subsubsection{Realistic dataset}
The Realistic datasets have been collected in 2019 from nine open-source software systems (i.e., ActiveMQ, Camel, Derby, Groovy, HBase, Hive,
JRuby, Lucene, and Wicket), which vary in size, domain, and defect ratio in order to reduce potential conclusion bias.
The data has been extracted from the JIRA Issue Tracking System of these software by following a rigorous procedure as explained elsewhere \cite{yatish2019mining}, resulting in less erroneous defect counts and hence representing a more realistic scenario of defective module collection. The metrics extracted include code, process, and ownership metrics for a total of 65 metrics (i.e., 54 code metrics, 5 process metrics, and 6 ownership metrics) as detailed in the original paper \cite{yatish2019mining}. 
In our experiment we consider two releases for each of these nine software systems, as listed in Table \ref{tab:stat-rep-datasets}. 
This allows us to explore the applicability of OCSVM for the cross-version defect prediction (CVDP) scenario, where data from one release is used to build prediction models to identify defects in subsequent releases. 
We also explore the cross-project defect prediction (CPDP) scenario by exploiting the Realistic datasets, where data from various software projects is used all together to build prediction models for predicting defective instances in a different target project. CPDP is useful when the target project lacks historical or sufficient local data \cite{herbold2017comparative}. However, it has been shown that CPDP is a more difficult prediction problem than CVDP due to the use of heterogeneous data \cite{nam2017heterogeneous}.



\subsection{Evaluation Criteria}
\label{sec:evalMeas}
The performance of a classification model is normally evaluated based on a confusion matrix 
describing four types of instances: True Positives (TP), defective modules correctly classified as defective; False Positives (FP), non-defective modules falsely classified as defective; False Negatives (FN), defective modules falsely classified as non-defective;
True Negatives (TN), defective modules correctly classified as defective.

The values in the confusion matrix are used to compute a set of evaluation measures. Common ones include Recall (which describes the proportion of defective modules that are actually classified as defective), Precision (which measures the proportion of modules that are actually defective out of the ones classified as defective), and F-Measure (which is the harmonic mean of Precision and Recall). However, when the data is imbalanced, which is frequently the case in defect prediction, 
it is recommended to use the Matthews Correlation Coefficient (MCC) as, unlike the other measures, it is a balanced measure and less prone to data imbalance bias \cite{moussa2022issta,shepperd2014researcher, yao2020assessing}.
Thus, in our empirical study we use MCC. 
MCC represents the correlation coefficient between the actual and predicted classifications:

$MCC = \frac{ TP \times TN - FP \times FN } {\sqrt{ (TP + FP) ( TP + FN ) ( TN + FP ) ( TN + FN ) } }$

\noindent it outputs a value between $-1$ and $+1$ where a value of $+1$ indicates a perfect prediction, a value of $0$ signifies that the prediction is no better than random guessing, and $-1$ represents a completely mis-classified output. 

We also investigate whether there is any statistical significance between the results obtained by the models, by using the Wilcoxon Signed-Rank Test \cite{woolson2007wilcoxon} with confidence limit $\alpha$=0.05 and Bonferroni correction ($\alpha/K$, where K is the number of hypotheses) for multiple statistical testing (the most conservatively cautious of all corrections) \cite{sarro2016multi}.
In addition, we check the effect size is worthy of interest by using the Vargha and Delaney's non-parametric effect size measure 
 $\hat{A}_{12}$, since it is recommended to use a standardised measure rather than a pooled one like the Cohen's $d$ when not all samples are normally distributed \cite{ArcuriB14}, as in our case. 
The $\hat{A}_{12}$ statistic measures the probability that an algorithm $A$ yields better values for a given performance measure $M$ than running another algorithm $B$, based on the following formula $\hat{A}_{12} = (R_1/m - (m + 1)/2)/n$, where $R_1$ is the rank sum of the first data group we are comparing, and $m$ and $n$ are the number of observations in the first and second data sample, respectively. If the two algorithms are equivalent, then $\hat{A}_{12} = 0.5$. Given the first algorithm performing better than the second, $\hat{A}_{12}$ is considered small for $0.6 \leq \hat{A}_{12} < 0.7$, medium for 0.7 $< \hat{A}_{12} < 0.8$, and large for $\hat{A}_{12} \geq 0.8$, although these thresholds are somewhat arbitrary  \cite{sarro2016multi}. 

\subsection{Validation}
\label{sec:design-eval-val}

For the within-project scenario experiments, involving the NASA data, we follow common practice 
for Hold-Out validation using 80\% of the data for training and the other 20\% for testing, and repeating this process 30 times, each time using a different seed, in order to reduce any possible bias resulting from the validation splits \cite{arcuri2014hitchhiker}.

For the experiments involving the Realistic data, we explore the performance of the models in two additional scenarios (namely CVDP and CVDP) given that this data consists of multiple releases as explained in Section \ref{sec:data}.
In the CVDP scenario, for each of the software systems, we train on one release and test on a different one, i.e., we train on version \textit{v$_{x}$} and test on version \textit{v$_{y}$}, where $x < y$ as done in previous work (see e.g., \cite{harman2014less}). 
In the CVDP, for each of the software systems, we consider the version with the higher release number as the test set and train the model on the union of the versions of the other datasets with a lower release number.
The versions used as train and test sets are not subsequent releases nor are they the system's most recent ones. In addition, there is always a window of at least five months between these releases. This reduces the likelihood of the snoring effect or unrealistic labelling as described in previous studies \cite{ahluwalia2020need,jimenez2019importance, bangash2020time}.


\subsection{Techniques}
\label{sec:Techniques}


SVM \cite{burges1998tutorial} is a classification technique based on the use of hyperplanes able to separate data points into two categories. Since there might be several hyperplanes that can correctly separate the data, SVM seeks to find the hyperplane that has the largest margin, in order to achieve a maximum separation between the two categories. When the data is not linearly separable, SVM does the mapping from input space to feature space. To achieve this, a kernel function is used instead of an inner product. This allows the formation of a non-linear decision boundary. OCSVM is an unsupervised version of SVM, whereby the technique trains on one class label only instead of two. Similar to its two-class counterpart SVM, it aims to draw a boundary around the instances that belong to the same class. However, given that this technique learns from one class only, it creates boundaries for the instances that belong to that class. Any instance that is not mapped inside the created boundary is considered an anomaly or an outlier, and hence classified as the other class.
In our empirical study we benchmark OCSVM with respect to a Random Classifier (RQ1) and four traditional and widely used two-classes classifiers: SVM \cite{burges1998tutorial} (RQ2), Logistic Regression \cite{lavalley2008logistic}, Naïve Bayes (NB) \cite{witten2005practical} and Random Forest (RF) \cite{breiman2001random} (RQ3).
We experiment with both tuned and non-tuned versions of SVM and a tuned version of RF since recent work has emphasized on the importance of hyperparameter tuning for defect prediction \cite{di2011genetic,tantithamthavorn2018impact, tantithamthavorn2016automated}.
We use the default parameters of {\tt Scikit Learn} \cite{scikitLearn} for the non-tuned version and perform Grid Search for the tuned ones (we indicate with \textit{technique$_{T}$} that hyperparameter tuning has been applied to a given techqniue) by using {\tt GridSearchCV} from the {\tt Scikit-Learn} v.0.20.2 library \cite{scikitLearn} in Python v.3.6.8. By applying hyperparameter tuning, we search for and obtain the best parameter values for the ML techniques used in our study based on MCC.
Since {\tt GridSearchCV} cannot be applied for classifiers that only learn from modules of the same class like OCSVM, we implement our own Grid Search method, namely {\tt GridSearchCV-OCSVM}. In the latter, we introduce instances of the other class (i.e., defective modules) in the testing fold of Grid Search's inner CV (which is only used to assess the hyper-parameters and not used in the training of the model or its validation). This allows us to obtain a confusion matrix, like the one obtained when {\tt GridSearchCV} is performed on a two-class predictor, from which we are able to calculate the MCC and choose the most suitable parameter values. The number of defective instances added to the testing fold reflects the proportion of classes in the original training set and are drawn randomly, with no duplicates.

\section{Empirical Study Results}
\label{sec:empiricalStudy}
In this section we report the results of our empirical study answering RQs1--3 for each of the scenarios investigated.

\subsection{Results for Within-Project Scenario}
\label{sec:concep-design-nasa}

\begin{table*}[tb]
\centering
\resizebox{0.65\linewidth}{!}{
\begin{tabular}{llrrrrrrrr}
\hline
 \multicolumn{1}{l}{\textbf{Validation}} & \multicolumn{1}{l}{\textbf{Dataset}}  & \multicolumn{1}{l}{\textbf{Random}} & \multicolumn{1}{l}{\textbf{OCSVM}} & \multicolumn{1}{l}{\textbf{SVM}} & \multicolumn{1}{l}{\textbf{OCSVM$_{T}$}} & \multicolumn{1}{l}{\textbf{SVM$_{T}$}} & \multicolumn{1}{l}{\textbf{NB}} & \multicolumn{1}{l}{\textbf{LR}} & \multicolumn{1}{l}{\textbf{RF$_{T}$}} \\ \hline
\multirow{6}{*}{\textbf{HoldOut}} & \textbf{CM1}      & 0.00                               & 0.00                               & 0.00                             & 0.09                                & 0.11                              & \textbf{0.17}                            & 0.13                            & -0.01                            \\
                                  & \textbf{MW1}      & 0.00                               & 0.00                               & 0.00                             & 0.11                                & -0.03                             & \textbf{0.32}                            & 0.24                            & 0.14                             \\
                                  & \textbf{KC3}      & -0.03                              & 0.00                               & 0.00                             & 0.24                                & -0.02                             & \textbf{0.32}                            & 0.19                            & 0.09                             \\
                                  & \textbf{PC1}      & 0.00                               & 0.00                               & 0.00                             & 0.12                                & 0.02                              & 0.26                            & 0.22                            & \textbf{0.31}                             \\
                                  & \textbf{PC3}      & 0.00                               & 0.00                               & 0.00                             & 0.15                                & 0.12                              & -0.03                           & 0.17                            & \textbf{0.20}                             \\
                                  & \textbf{PC5}      & 0.00                               & 0.11                               & -0.04                            & \textbf{0.29}                                & 0.27                              & -0.02                           & 0.25                            & 0.25                             \\ \hline
\multirow{9}{*}{\textbf{CVDP}}    & \textbf{ActiveMQ} & 0.00                               & 0.03                               & 0.00                             & 0.18                                & 0.27                              & 0.27                            & 0.26                            & \textbf{0.35}                             \\
                                  & \textbf{Camel}    & 0.00                               & 0.11                               & 0.00                             & 0.17                                & 0.14                              & \textbf{0.19}                            & 0.00                            & 0.15                             \\
                                  & \textbf{Derby}    & 0.00                               & 0.10                               & 0.06                             & 0.27                                & 0.37                              & 0.08                            & 0.13                            & \textbf{0.40}                            \\
                                  & \textbf{Groovy}   & 0.00                               & 0.18                               & 0.04                             & 0.20                                & 0.26                              & -0.07                           & 0.07                            & \textbf{0.30}                             \\
                                  & \textbf{Hbase}    & 0.00                               & 0.15                               & -0.06                            & 0.25                                & 0.19                              & 0.05                            & 0.03                            & \textbf{0.29}                             \\
                                  & \textbf{Hive}     & 0.00                               & 0.12                               & -0.01                            & 0.20                                & 0.17                              & 0.00                            & 0.02                            & \textbf{0.27}                             \\
                                  & \textbf{JRuby}    & 0.01                               & 0.10                               & 0.00                             & 0.19                                & 0.07                              & 0.17                            & 0.07                            & \textbf{0.34}                             \\
                                  & \textbf{Lucene}   & 0.00                               & 0.03                               & 0.00                             & \textbf{0.10}                                & 0.08                              & 0.14                            & 0.05                            & 0.07                             \\
                                  & \textbf{Wicket}   & 0.00                               & 0.01                               & 0.00                             & 0.03                                & 0.06                              & \textbf{0.20}                            & -0.01                           & 0.16                             \\ \hline
         \multirow{9}{*}{\textbf{CPDP}} & \textbf{ActiveMQ} & 0.00  & 0.05 & -0.03 & \textbf{0.25} & 0.21 & 0.00 & 0.08  & 0.22 \\
                      & \textbf{Camel}    & 0.00  & 0.05 & 0.00  & 0.17 & 0.10 & 0.00 & 0.01  & \textbf{0.19} \\
                      & \textbf{Derby}    & 0.01  & 0.03 & 0.00  & \textbf{0.28} & 0.13 & 0.11 & 0.01  & 0.14 \\
                      & \textbf{Groovy}   & -0.01 & 0.12 & 0.00  & 0.23 & 0.07 & 0.07 & -0.01 & \textbf{0.25} \\
                      & \textbf{HBase}    & 0.01  & 0.09 & 0.00  & 0.26 & 0.20 & 0.03 & -0.04 & \textbf{0.29} \\
                      & \textbf{Hive}     & -0.01 & 0.10  & 0.00  & \textbf{0.17} & 0.08 & 0.02 & 0.04  & 0.14 \\
                      & \textbf{JRuby}    & 0.00  & 0.04 & 0.00  & 0.20  & 0.07 & 0.18 & 0.17  & \textbf{0.33} \\
                      & \textbf{Lucene}   & 0.00  & 0.06 & 0.00  & 0.12 & 0.09 & 0.06 & 0.01  & \textbf{0.18} \\
                      & \textbf{Wicket}   & 0.00  & 0.04 & 0.00  & \textbf{0.19} & 0.12 & 0.00 & 0.04  & 0.17 \\ \hline                         
\end{tabular}
}
\caption{\textbf{RQs1-3.} Average MCC values achieved by each of the techniques over 30 runs.} 
\label{tab:rep-study-results}
\end{table*}

\textbf{RQ1. OCSVM vs. Random: } 
Table \ref{tab:rep-study-results} shows that the MCC values obtained by OCSVM are higher than those obtained by Random on two datasets (i.e., KC3 and PC5), while on the other four datasets, the former obtains the same results, with values indicating that it is no better than random guessing. The Wilcoxon test and $\hat{A}_{12}$ effect size measures, reported in Table \ref{tab:rq1-rq2-wilc-rep}, also support this by showing that the difference achieved on PC5 is statistically significant. On the other hand, OCSVM$_{T}$ outperforms Random on all datasets, with a statistically significant difference and a large effect size on four datasets and medium effect size on the remaining two.

\textbf{RQ2. OCSVM vs. SVM:} To address this question, we compare both OCSVM and OCSVM$_{T}$ with their two-class couterparts, SVM and SVM$_{T}$. It is clear from the results shown in Table \ref{tab:rep-study-results} that hyperparameter tuning enhances the performance of these techniques, as OCSVM$_{T}$ and SVM$_{T}$ generally obtain better results than OCSM and SVM, respectively. We can also observe that, while SVM and OCSVM obtain the same results on five datasets, with the latter performing better on the remaining dataset (i.e., PC5), OCSVM$_{T}$ outperforms both SVM and OCSVM, on all the datasets, with a statistically significant difference, with five of them showing a large effect size and the remaining dataset having a medium one, as shown in Table \ref{tab:rq1-rq2-wilc-rep}. OCSVM$_{T}$ also outperforms SVM$_{T}$ 
on five out of six datasets, with the difference being statistically significant on three of them with a large effect size.

\begin{table*}[]
\centering
\resizebox{0.9\linewidth}{!}{
\begin{tabular}{llrr|rrrrr}
\hline
Validation & Dataset & OCSVM vs. Random  & OCSVM$_{T}$ vs. Random & OCSVM vs. SVM & OCSVM vs. SVM$_{T}$ & OCSVM$_{T}$ vs. SVM & OCSVM$_{T}$ vs. SVM$_{T}$ \\ \hline
\multirow{6}{*}{HoldOut} & CM1      & 0.484 (0.53) & \textbf{0.003 (0.71)} & 0.373 (0.50) & 1.000 (0.00)        & \textbf{\textless0.001 (0.90)}  & 0.786 (0.47) \\
                         & KC3      & 0.245 (0.57) & \textbf{\textless0.001 (0.94)}     & 0.164 (0.50) & \textbf{\textless0.001 (1.00)}        & \textbf{\textless0.001 (1.00)}    & \textbf{\textless0.001 (1.00)}        \\
                         & MW1      & 0.452 (0.53) & \textbf{0.003 (0.72)} & 0.774 (0.50) & 0.050 (0.68)  & \textbf{\textless0.001 (0.77)} & \textbf{\textless0.001 (0.80)}      \\
                         & PC1      & 0.580 (0.47)  & \textbf{\textless0.001 (0.89)}     & \textbf{0.009 (0.60)} & 0.755 (0.48) & \textbf{\textless0.001 (0.97)} & \textbf{\textless0.001 (0.86)}     \\
                         & PC3      & 0.540 (0.50)   & \textbf{\textless0.001 (0.97)}     & 0.886 (0.50) & 1.000 (0.08)     & \textbf{\textless0.001 (0.97)} & 0.060 (0.65)  \\
                         & PC5      & \textbf{0.021 (0.64)} & \textbf{\textless0.001 (0.81)}     & \textbf{\textless0.001 (0.92)}    & 1.000 (0.00)        & \textbf{\textless0.001 (0.92)} & 0.180 (0.60)   \\ \hline
\multirow{9}{*}{CVDP}    & ActiveMQ & \textbf{\textless0.001 (0.93)}     & \textbf{\textless0.001 (1.00)}        & \textbf{\textless0.001 (1.00)}       & 1.000 (0.00)        & \textbf{\textless0.001 (1.00)}    & 1.000 (0.00)        \\
                         & Camel    & \textbf{\textless0.001 (1.00)}        & \textbf{\textless0.001 (1.00)}        & \textbf{\textless0.001 (1.00)}       & 1.000 (0.00)        & \textbf{\textless0.001 (1.00)}    & \textbf{\textless0.001 (1.00)}        \\
                         & Derby    & \textbf{\textless0.001 (1.00)}        & \textbf{\textless0.001 (1.00)}        & \textbf{\textless0.001 (1.00)}       & 1.000 (0.00)        & \textbf{\textless0.001 (1.00)}    & 1.000 (0.00)        \\
                         & Groovy   & \textbf{\textless0.001 (1.00)}        & \textbf{\textless0.001 (1.00)}        & \textbf{\textless0.001 (1.00)}       & 1.000 (0.00)        & \textbf{\textless0.001 (1.00)}    & 1.000 (0.00)        \\
                         & HBase    & \textbf{\textless0.001 (1.00)}        & \textbf{\textless0.001 (1.00)}        & \textbf{\textless0.001 (1.00)}       & 1.000 (0.00)        & \textbf{\textless0.001 (1.00)}    & \textbf{\textless0.001 (1.00)}        \\
                         & Hive     & \textbf{\textless0.001 (1.00)}        & \textbf{\textless0.001 (1.00)}        & \textbf{\textless0.001 (1.00)}       & 1.000 (0.00)        & \textbf{\textless0.001 (1.00)}    & \textbf{\textless0.001 (0.90)}      \\
                         & JRuby    & \textbf{\textless0.001 (1.00)}        & \textbf{\textless0.001 (1.00)}        & \textbf{\textless0.001 (1.00)}       & \textbf{\textless0.001 (1.00)}        & \textbf{\textless0.001 (1.00)}    & \textbf{\textless0.001 (1.00)}        \\
                         & Lucene   & \textbf{\textless0.001 (0.90)}      & \textbf{\textless0.001 (1.00)}        & \textbf{\textless0.001 (1.00)}       & 1.000 (0.00)        & \textbf{\textless0.001 (1.00)}    & \textbf{\textless0.001 (1.00)}        \\
                         & Wicket   & \textbf{\textless0.001 (0.73)}     & \textbf{\textless0.001 (0.93)}     & \textbf{\textless0.001 (1.00)}       & 1.000 (0.00)        & \textbf{\textless0.001 (0.93)} & 1.000 (0.00)        \\ \hline
\multirow{9}{*}{CPDP}    & ActiveMQ & \textbf{\textless0.001 (1.00)}        & \textbf{\textless0.001 (1.00)}        & \textbf{\textless0.001 (1.00)}       & 1.000 (0.00)        & \textbf{\textless0.001 (1.00)}    & \textbf{\textless0.001 (1.00)}        \\
                         & Camel    & \textbf{\textless0.001 (1.00)}        & \textbf{\textless0.001 (1.00)}        & \textbf{\textless0.001 (1.00)}       & 1.000 (0.00)        & \textbf{\textless0.001 (1.00)}    & \textbf{\textless0.001 (1.00)}        \\
                         & Derby    & \textbf{\textless0.001 (0.90)}      & \textbf{\textless0.001 (1.00)}        & \textbf{\textless0.001 (1.00)}       & 1.000 (0.00)        & \textbf{\textless0.001 (1.00)}    & \textbf{\textless0.001 (1.00)}        \\
                         & Groovy   & \textbf{\textless0.001 (1.00)}        & \textbf{\textless0.001 (1.00)}        & \textbf{\textless0.001 (1.00)}       & \textbf{\textless0.001 (1.00)}        & \textbf{\textless0.001 (1.00)}    & \textbf{\textless0.001 (1.00)}        \\
                         & HBase    & \textbf{\textless0.001 (1.00)}        & \textbf{\textless0.001 (1.00)}        & \textbf{\textless0.001 (1.00)}       & 1.000 (0.00)        & \textbf{\textless0.001 (1.00)}    & \textbf{\textless0.001 (1.00)}        \\
                         & Hive     & \textbf{\textless0.001 (1.00)}        & \textbf{\textless0.001 (1.00)}        & \textbf{\textless0.001 (1.00)}       & \textbf{\textless0.001 (1.00)}        & \textbf{\textless0.001 (1.00)}    & \textbf{\textless0.001 (1.00)}        \\
                         & JRuby    & \textbf{\textless0.001 (0.93)}     & \textbf{\textless0.001 (1.00)}        & \textbf{\textless0.001 (1.00)}       & 1.000 (0.00)        & \textbf{\textless0.001 (1.00)}    & \textbf{\textless0.001 (1.00)}        \\
                         & Lucene   & \textbf{\textless0.001 (1.00)}        & \textbf{\textless0.001 (1.00)}        & \textbf{\textless0.001 (1.00)}       & 1.000 (0.00)        & \textbf{\textless0.001 (1.00)}    & \textbf{\textless0.001 (1.00)}        \\
                         & Wicket   & \textbf{\textless0.001 (0.97)}     & \textbf{\textless0.001 (1.00)}        & \textbf{\textless0.001 (1.00)}       & 1.000 (0.00)        & \textbf{\textless0.001 (1.00)}    & \textbf{\textless0.001 (1.00)}        \\ \hline
\end{tabular}}
\caption{\textbf{RQs1-2.} Wilcoxon test and $\hat{A}_{12}$ effect size (p-value($\hat{A}_{12}$)) results for OCSVM vs. Random and SVM.}
\label{tab:rq1-rq2-wilc-rep}
\end{table*}

\textbf{RQ3. OCSVM vs. traditional ML:} To answer this question, we compare the performance of OCSVM to that of NB, LR and RF$_{T}$. We can observe from Table \ref{tab:rep-study-results} and \ref{tab:rq3-wilc-rep} that OCSVM does not outperform LR, only preforms better than RF$_{T}$ on one dataset with a statistical significance and a large effect size and outperforms NB on two datasets with the differences being statistically significant and having a medium effect size. On the other hand, OCSVM$_{T}$ outperforms traditional ML techniques on seven out of the 18 cases studied. Specifically, it outperforms NB and LR on two datasets each, obtaining differences that are statistically significant with large effect sizes when compared to NB. OCSVM$_{T}$ also outperforms RF$_{T}$ on three datasets with two of them being statistically significant and having a large effect size.
\begin{table*}[tb]
\centering
\resizebox{0.9\linewidth}{!}{
\begin{tabular}{llrrr|rrr}
\hline
 Validation & Dataset  & OCSVM vs. NB & OCSVM vs. LR & OCSVM vs. RF$_{T}$ & OCSVM$_{T}$ vs. NB & OCSVM$_{T}$ vs. LR & OCSVM$_{T}$ vs. RF$_{T}$ \\ \hline
\multirow{6}{*}{HoldOut} & CM1      & 1.000 (0.15)     & 0.999 (0.37) & \textbf{\textless0.001 (0.80)}      & 0.989 (0.34) & 0.755 (0.45) & \textbf{\textless0.001 (0.92)}     \\
                         & KC3      & 1.000 (0.13)     & 1.000 (0.22)     & 1.000 (0.00)        & 0.962 (0.34) & 0.067 (0.61) & \textbf{\textless0.001 (0.94})     \\
                         & MW1      & 1.000 (0.03)     & 1.000 (0.33)     & 0.999 (0.33) & 1.000 (0.17)     & 0.975 (0.40)  & 0.680 (0.47)  \\
                         & PC1      & 1.000 (0.07)     & 1.000 (0.20)      & 1.000 (0.07)     & 1.000 (0.20)      & 0.985 (0.34) & 1.000 (0.15)     \\
                         & PC3      & \textbf{0.005 (0.77)} & 1.000 (0.27)     & 1.000 (0.00)        & \textbf{\textless0.001 (0.95)}     & 0.604 (0.46) & 0.974 (0.32) \\
                         & PC5      & \textbf{0.008 (0.64)} & 0.995 (0.35) & 1.000 (0.04)     & \textbf{\textless0.001 (0.87)}     & 0.271 (0.54) & 0.089 (0.63) \\ \hline
\multirow{9}{*}{CVDP}    & ActiveMQ & 1.000 (0.00)        & 1.000 (0.00)        & 1.000 (0.00)        & 1.000 (0.17)     & 1.000 (0.17)     & 1.000 (0.00)        \\
                         & Camel    & 1.000 (0.00)        & \textbf{\textless0.001 (1.00)}        & 1.000 (0.00)        & 1.000 (0.00)        & \textbf{\textless0.001 (1.00)}        & \textbf{\textless0.001 (0.99)}     \\
                         & Derby    & \textbf{\textless0.001 (1.00)}        & 1.000 (0.00)        & 1.000 (0.00)        & \textbf{\textless0.001 (1.00)}        & \textbf{\textless0.001 (1.00)}        & 1.000 (0.00)        \\
                         & Groovy   & \textbf{\textless0.001 (1.00)}        & \textbf{\textless0.001 (1.00)}        & 1.000 (0.00)        & \textbf{\textless0.001 (1.00)}        & \textbf{\textless0.001 (1.00)}        & 1.000 (0.01)     \\
                         & HBase    & \textbf{\textless0.001 (1.00)}        & \textbf{\textless0.001 (1.00)}        & 1.000 (0.00)        & \textbf{\textless0.001 (1.00)}        & \textbf{\textless0.001 (1.00)}        & 1.000 (0.00)        \\
                         & Hive     & \textbf{\textless0.001 (1.00)}        & \textbf{\textless0.001 (1.00)}        & 1.000 (0.00)        & \textbf{\textless0.001 (1.00)}        & \textbf{\textless0.001 (1.00)}        & 1.000 (0.00)        \\
                         & JRuby    & 1.000 (0.00)        & \textbf{\textless0.001 (1.00)}        & 1.000 (0.00)        & \textbf{\textless0.001 (0.97)}     & \textbf{\textless0.001 (1.00)}        & 1.000 (0.00)        \\
                         & Lucene   & 1.000 (0.00)        & 1.000 (0.00)        & 1.000 (0.03)     & 1.000 (0.00)        & \textbf{\textless0.001 (1.00)}        & \textbf{\textless0.001 (0.90)}      \\
                         & Wicket   & 1.000 (0.00)        & \textbf{\textless0.001 (1.00)}        & 1.000 (0.00)        & 1.000 (0.00)        & \textbf{\textless0.001 (0.93)}     & 1.000 (0.00)        \\ \hline
\multirow{9}{*}{CPDP}    & ActiveMQ & \textbf{\textless0.001 (1.00)}        & 1.000 (0.00)        & 1.000 (0.00)        & \textbf{\textless0.001 (1.00)}        & \textbf{\textless0.001 (1.00)}        & \textbf{\textless0.001 (1.00)}        \\
                         & Camel    & \textbf{\textless0.001 (1.00)}        & \textbf{\textless0.001 (1.00)}        & 1.000 (0.00)        & \textbf{\textless0.001 (1.00)}        & \textbf{\textless0.001 (1.00)}        & 1.000 (0.20)      \\
                         & Derby    & 1.000 (0.00)        & \textbf{\textless0.001 (1.00)}        & 1.000 (0.00)        & \textbf{\textless0.001 (1.00)}        & \textbf{\textless0.001 (1.00)}        & \textbf{\textless0.001 (1.00)}        \\
                         & Groovy   & \textbf{\textless0.001 (1.00)}        & \textbf{\textless0.001 (1.00)}        & 1.000 (0.00)        & \textbf{\textless0.001 (1.00)}        & \textbf{\textless0.001 (1.00)}        & 1.000 (0.13)     \\
                         & HBase    & \textbf{\textless0.001 (1.00)}        & \textbf{\textless0.001 (1.00)}        & 1.000 (0.00)        & \textbf{\textless0.001 (1.00)}        & \textbf{\textless0.001 (1.00)}       & 1.000 (0.05)     \\
                         & Hive     & \textbf{\textless0.001 (1.00)}          & \textbf{\textless0.001 (1.00)}          & 1.000 (0.00)        & \textbf{\textless0.001 (1.00)}         & \textbf{\textless0.001 (1.00)}          & \textbf{\textless0.001 (1.00)}          \\
                         & JRuby    & 1.000 (0.00)        & 1.000 (0.00)        & 1.000 (0.00)        & \textbf{\textless0.001 (1.00)}          & \textbf{\textless0.001 (1.00)}          & 1.000 (0.00)        \\
                         & Lucene   & 1.000 (0.00)        & \textbf{\textless0.001 (1.00)}          & 1.000 (0.00)        & \textbf{\textless0.001 (1.00)}          & \textbf{\textless0.001 (1.00)}          & 1.000 (0.00)        \\
                         & Wicket   & \textbf{\textless0.001 (1.00)}          & \textbf{\textless0.001 (1.00)}          & 1.000 (0.00)        & \textbf{\textless0.001 (1.00)}          & \textbf{\textless0.001 (1.00)}          & \textbf{\textless0.001 (0.93)}     \\ \hline
\end{tabular}}
\caption{\textbf{RQ3.} Wilcoxon test and $\hat{A}_{12}$ effect size (p-value($\hat{A}_{12}$)) results for OCSVM vs. traditional ML.} 
\label{tab:rq3-wilc-rep}
\end{table*}
Therefore, based on the above results we can state that:

\finding{
		OCSVM performs similarly to Random, while OCSVM$_{T}$ statistically significantly outperforms it. When compared to its two-class counterpart SVM, OCSVM performs similarly. Whereas, OCSVM$_{T}$ obtains better results than SVM$_{T}$ on five out of the six datasets studied with three of the differences being statistically significant. However, OCSVM and OCSVM$_{T}$ outperform the traditional ML techniques in three and seven of the 18 cases studied, respectively.}

\subsection{Results for Cross-Version Scenario}
\label{sec:concep-design-realistic}




\textbf{RQ1. OCSVM vs. Random:} Table \ref{tab:rep-study-results} shows the MCC values obtained by OCSVM and Random. We can observe that OCSVM obtains better results on all datasets, with the difference always being statistically significant and the effect size being large for eight of these nine datasets. When comparing OCSVM$_{T}$ and Random, results show that OCSVM$_{T}$ also outperforms Random on all datasets. The Wilcoxon Test and $\hat{A}_{12}$ effect size results, described in Table \ref{tab:rq1-rq2-wilc-rep}, support this conclusion as they show a statistically significant difference and a large effect size on all cases considered. 

\textbf{RQ2. OCSVM vs. SVM:} To address RQ2, we compare the performance of OCSVM with that of SVM. Results in Table \ref{tab:rep-study-results} indicate that 
OCSVM always outperforms its two-class counterpart SVM. Table \ref{tab:rq1-rq2-wilc-rep} also indicates that this conclusion is supported by the statistical tests, as it shows that the difference is always statistically significant with the effect size being large. However, this is not always the case when comparing the hyperparameter tuned version of both of these models, OCSVM$_{T}$ and SVM$_{T}$. Results show that the former obtains better results than the latter on five out of the nine cases studied with all differences being statistically significant and the effect size being large. 


\textbf{RQ3. OCSVM vs. traditional ML:}
Based on the results reported in Tables \ref{tab:rep-study-results} and \ref{tab:rq3-wilc-rep}, we can observe that RF$_{T}$ generally performs better than the other techniques, achieving the highest MCC values in six out of the nine cases under study. However, when analysing the results obtained by NB, LR, and OCSVM we can see that the latter performs better than LR on 67\% of the cases (i.e., six out of nine) and better than NB on 44\% (four out of nine cases) respectively, with the differences always being statistically significant and having a large effect size. 
While when hyperparameter tuning is applied, OCSVM$_{T}$ obtains better results than LR, NB and RF$_{T}$ 89\% of the time (eight out of nine cases),  56\% of the time (i.e., five out of nine cases) and 22\% of the time (two out of nine cases) with the differences always being statistically significant and having large effect sizes. 

\finding{Both OCSVM and OCSVM$_{T}$ significantly outperform Random, so passing our sanity check; OCSVM also performs significantly better than SVM, however whenever tuning is applied to SVM (SVM$_{T}$), the latter achieves the best results. Both OCSVM and OCSVM$_{T}$ are significantly better than NB, LR and RF$_{T}$ in 50\%, 72\% and 22\% of the cases, respectively.} 

\subsection{Results for the Cross-Project Scenario}
\textbf{RQ1. OCSVM vs. Random: }
To address RQ1, we compare the performance  of  OCSVM with that of a Random classifer. Results reported in Table \ref{tab:rep-study-results} show that both OCSVM and OCSVM$_{T}$ outperform Random on all datasets with all differences being statistically significant and the effect size always being large as shown in Table \ref{tab:rq1-rq2-wilc-rep}. 

\textbf{RQ2. OCSVM vs. SVM:}
When compared to its two-class counterpart, OCSVM performs better than SVM on all the cases considered (see Table \ref{tab:rep-study-results}). The Wilcoxon and $\hat{A}_{12}$ results, reported in Table \ref{tab:rq1-rq2-wilc-rep}, confirm that these differences are statistically significant with a large effect size. When hyperparameter tuning is applied, OCSVM$_{T}$  
also outperforms both SVM and SVM$_{T}$ on all nine datasets with statistically significant differences and large effect size. However, the MCC values as well as the Wilcoxon and $\hat{A}_{12}$ results show that SVM$_{T}$ outperforms OCSVM in seven out of the nine cases considered and performs better than SVM in all cases studied.

\textbf{RQ3. OCSVM vs. traditional ML:}
To investigate RQ3, we compare the performance of OCSVM and OCSVM$_{T}$ to that of the traditional two-class classifiers (i.e., NB, LR and RF). 
Results reported in Table \ref{tab:rep-study-results} show that both OCSVM$_{T}$ and RF$_{T}$ perform better than the other techniques. Specifically, while OCSVM performs better than NB and LR on 67\% of the cases (six out of nine cases each) 
, it always performs worse than RF$_{T}$. On the other hand, when hyperparameter tuning is applied, OCSVM$_{T}$ always performs better than NB and LR with the differences being statistically significant and the effect size being large in all cases. When compared to RF$_{T}$, OCSVM$_{T}$ performs better in 44\% of the cases (four out of nine) with the differences always being statistically significant and the effect size being large.


\finding{Both OCSVM and OCSVM$_{T}$ significantly outperform Random and SVM. While OCSVM generally outperforms NB and LR in most of the cases, it never performs better than RF$_{T}$. However, when tuning is applied, OCSVM$_{T}$ always outperforms NB and LR in all cases and it competes with RF$_{T}$, performing better than the latter in 44\% of the cases.}
		 

\section{Replication}
\label{sec:ExactRep}
The results of our empirical study (Section \ref{sec:empiricalStudy}) show that OCSVM generally outperforms its two-class counterpart, SVM, however it does not achieve results consistently higher than traditional two-class classifiers and its performance is not as promising as shown in the work of \originalp. For due diligence and completeness, we replicate their study \originalcite. 
Below we describe the design and results of our replication, a summary of the design is given in Table \ref{tab:studies2}. 


\begin{table*}[tb]
\centering
 \resizebox{1\linewidth}{!}{
\begin{tabular}{l|ll} 
\hline
 & Original Study  & Replication  \\ \hline
\multirow{1}{*}{Validation Approach} & \begin{tabular}[c]{@{}l@{}}Hold-out: 10\% for training,  90\% for testing, 20 repetitions \end{tabular}  &  \begin{tabular}[c]{@{}l@{}}Hold-out: 10\% for training,  90\% for testing, 30 repetitions\end{tabular} \\ \hline
Techniques   & \begin{tabular}[c]{@{}l@{}} OCSVM, NB, NB Log filter. Random under-sampling boosting,  Cost sensitive SVM\end{tabular}  &  \begin{tabular}[c]{@{}l@{}} OCSVM,  NB,  LR,  RF,  SVM,  Random\end{tabular}   \\ \hline
\multirow{1}{*}{Evaluation Criteria} & \begin{tabular}[c]{@{}l@{}} Recall, False Positive Rate, G-Mean\end{tabular} & \begin{tabular}[c]{@{}l@{}}  MCC, G-Mean, Wilcoxon Signed Rank Test,  Vargha \& Delaney $\hat{A}_{12}$ \end{tabular} \\ \hline
Datasets & \begin{tabular}[c]{@{}l@{}} CM1,  KC3, MC1,  MW1,  PC1,  PC2\end{tabular}  & \begin{tabular}[c]{@{}l@{}} CM1, KC3, MC1,  MW1, PC1, PC2\end{tabular} \\ \hline
\end{tabular}}
\caption{Summary of the empirical design adopted in the original study and in our replication.}
\label{tab:studies2}
\end{table*}

\subsection{Design}

\subsubsection{Research Questions}
\label{sec:exact-design-rqs}
\originalp's study did not explicitly state multiple research questions, but rather their overall research goal to investigate whether OCSVM can be used to predict defects, and whether it would outperform other ML techniques. Hence, in this replication we aim to address the same goal organised as the research questions we described in Section \ref{sec:concep-design-rqs}. We therefore investigate whether (RQ1) OCSVM outperforms the Random Classifier; (RQ2) OCSVM outperforms its two-class counterpart, namely SVM; (RQ3) OCSVM outperforms traditional two-class classification techniques widely used for defect prediction. 

\subsubsection{Datasets}
\label{sec:exact-design-datasets}
\originalp \space investigate six highly imbalanced datasets obtained from the public NASA repository \cite{Sayyad-Shirabad+Menzies:2005}:  CM1, KC3, MC1, MW1, PC1, PC2. 
In our replication we use the same datasets available from the tera-PROMISE repository \cite{Sayyad-Shirabad+Menzies:2005}. 
We report in Table \ref{tab:stat-details-exact-rep} the number of modules and percentage of faulty modules per dataset. We observe that three out of the six datasets used (i.e., KC3, MW1, PC1) are identical, whereas the other three datasets (i.e., CM1, MC1, PC2) vary slightly in the number of instances and percentage of defects from those used by \originalp. \space
The reason for this difference cannot be determined given that no indication of pre-processing was stated in \originalp's study. 


\subsubsection{Validation and Evaluation Criteria} 
\label{sec:exact-rep-val}

\originalp  \space performed 20 independent Hold-Out validations, where each time, 10\% of the data was randomly selected for training and 90\% for testing. We, perform the same Hold-Out validation, but we increase the number of independent runs to 30 in order to gather more robust results. 
Recall, False Positive Rate and G-mean were used to evaluate and compare the performances of the techniques, however, the main conclusions were drawn based on G-Mean \originalcite. We decided to assess the performance of the techniques using MCC (see  Section \ref{sec:evalMeas}), since it is a robust evaluation measure \cite{shepperd2014researcher, yao2020assessing}. For completeness, we also include the G-mean results for OCSVM and compare them with those obtained in the original study (OCSVM-O) in Table \ref{tab:exact-rep-gmean}. We observe that the G-mean values obtained with OCSVM in our study are much lower than those reported in the work of \originalp. 
The original study did not perform any statistical analysis, while we use the Wilcoxon signed rank test and the Vargha and Delaney Â12 effect size (described in Section \ref{sec:evalMeas}) to check for any statistically significant difference in order to strengthen the robustness of our conclusions. 

\subsubsection{Techniques} 
\label{sec:exact-rep-tech}
To carry out a fair comparison with previous work, we compare OCSVM to SVM and RF, NB and LR using their non-tuned version (see Section \ref{sec:Techniques}) as the original work did not perform any tuning.


\begin{table}[]
\centering
\resizebox{0.75\linewidth}{!}{
\begin{tabular}{lrr}
\hline
      & Original Study           & Replication        \\ \cline{1-3} 
  Dataset   & Modules (faulty \%) & Modules (faulty \%) \\ \hline
  CM1 & 496 (9.68\%)  & 498 (9.83\%) \\
  KC3 & 458 (9.39\%)  & 458 (9.39\%)\\
  MC1 & 9277 (0.73\%) & 9466 (0.72\%)\\
  MW1 & 403 (7.69\%)  & 403 (7.69\%)\\
  PC1 & 1107 (6.87\%) & 1107 (6.87\%)\\
  PC2 & 5460 (0.42\%) & 5589 (0.41\%) \\ 
  \cline{1-3}
\end{tabular}
}
\caption{\textbf{Original Study and Replication Data:} Total number of modules and percentage of faulty modules used in the original and replication studies.}
\label{tab:stat-details-exact-rep}
\end{table}

\subsection{Results}

\begin{table}[]
\centering
\resizebox{0.75\linewidth}{!}{
\begin{tabular}{lrrrrrrr}
\hline
 Dataset & Random & OCSVM  & SVM  & RF            & NB            & LR   \\ \hline
\multirow{1}{*}{CM1}       & 0.03  & 0.03     & 0.01 & 0.08 & 0.11          & 0.13 \\ 
\multirow{1}{*}{KC3}      & 0.00  & 0.10      & 0.00 & 0.11 & 0.21          & 0.14 \\ 
\multirow{1}{*}{MC1}      & 0.00  & 0.07      & 0.24 & 0.23 & 0.13          & 0.09 \\ 
\multirow{1}{*}{MW1}      & 0.01  & 0.01      & 0.00 & 0.14 & 0.19          & 0.16 \\ 
\multirow{1}{*}{PC1}      & 0.00  & 0.00       & 0.12 & 0.19 & 0.15          & 0.15 \\
\multirow{1}{*}{PC2}      & 0.00  & 0.04      & 0.00 & 0.02 & 0.07          & 0.05 \\ \hline
\end{tabular}
}
\caption{\textbf{RQs1-3. Replication Study:} Average MCC values over 30 runs obtained by OCSVM, Random and traditional two-class classifiers on the six NASA datasets.}
\label{tab:replication}
\end{table}

\begin{table*}[]
\centering
\resizebox{0.75\linewidth}{!}{
\begin{tabular}{lrrrrr}
\hline
Dataset   & OCSVM vs. Random        & OCSVM vs. SVM          & OCSVM vs. RF           & OCSVM vs. NB           & OCSVM vs. LR           \\ \hline
CM1 & 0.306 (0.54)   & 0.007 (0.77) & 0.997 (0.26) & 0.994 (0.14) & 1.000 (0.13)     \\
KC3 & \textless0.001 (0.99)       & \textless0.001 (1.00)        & 0.846 (0.43) & 1.000 (0.14)     & 0.998 (0.27) \\
MC1 & \textless0.001 (1.00)          & 1.000 (0.21)     & 1.000 (0.23)     & 0.999 (0.14) & 0.396 (0.59) \\
MW1 & 0.245 (0.55)   & \textless0.001 (0.85)     & 1.000 (0.29)      & 1.000 (0.17)     & 1.000 (0.09)     \\
PC1 & 0.169 (0.63)   & 1.000 (0.31)     & 1.000 (0.06)     & 1.000 (0.06)     & 1.000 (0.07)     \\
PC2 & \textless0.001 (1.00)          & \textless0.001 (1.00)        & 0.002 (0.83) & 0.983 (0.37) & 0.708 (0.48) \\  \hline
\end{tabular}}
\caption{\textbf{RQs1-3. Replication Study:} Results of the Wilcoxon test and $\hat{A}_{12}$ effect size (p-value($\hat{A}_{12}$)) comparing OCSVM with each of the other techniques.}
\label{tab:wilc-replication}
\end{table*}

\begin{table}[tb]
\centering
\resizebox{0.50\linewidth}{!}{
\begin{tabular}{lrr}
\hline
\multicolumn{1}{r}{Dataset} & OCSVM & OCSVM-O \\ \hline
\multirow{1}{*}{CM1} & 0.11 & 0.63  \\
\multirow{1}{*}{KC3} & 0.30 & 0.66  \\
\multirow{1}{*}{MC1} & 0.62 & 0.84  \\
\multirow{1}{*}{MW1} & 0.04 & 0.61  \\
\multirow{1}{*}{PC1} & 0.15 & 0.64  \\
\multirow{1}{*}{PC2} & 0.60 & 0.76  \\ \hline
\end{tabular}
}
\caption{\textbf{Replication Study:} G-mean values obtained by one-class SVM in our replication (OCSVM) and the original study (OCSVM-O).}
\label{tab:exact-rep-gmean}
\end{table}

\textbf{RQ1. OCSVM vs. Random:}
To address RQ1, we compare the performance of OCSVM to that of a Random Classifier. From Tables \ref{tab:replication} and \ref{tab:wilc-replication} we can observe that 
OCSVM outperforms the Random classifier on three out of the six datasets with statistically significant difference and a large effect size. On the other three datasets, OCSVM obtains the same results as the random classifier. 

\textbf{RQ2. OCSVM vs. SVM:}
To answer RQ2, we compare the performance of OCSVM with its two-class counterpart SVM. By looking at the 
MCC values reported in Table \ref{tab:replication}, we can see that OCSVM 
performs similarly or better than SVM on four out of the six datasets (i.e., CM1, KC3, MW1, PC2) with differences being statistically significant on all four of them and the effect size being medium in one case and large in the three other cases. This shows that by learning from the non-defective modules only, a technique like SVM is able to perform similarly and sometimes better than when trained on both, defective and non-defective modules. 

\textbf{RQ3. OCSVM vs. traditional ML:}
In order to verify whether OCSVM outperforms traditional two-class classifiers, we compare its performance to that of three different techniques widely used in defect prediction studies (i.e., RF, NB, LR). Results, reported in Tables \ref{tab:replication} and \ref{tab:wilc-replication}, show that 
OCSVM only outperforms RF in one case, with the difference being statistically significant and the effect size being large. When compared to LR and NB, results show that OCSVM does not outperform these two techniques on any of the 12 cases considered. 

\finding{Our replication study shows that OCSVM statistically significantly outperforms the Random classifier on three out of the six cases considered. OCSVM also performs similarly or better than SVM, with statistical significant improvements in four out of 6 cases. However, OCSVM does not perform better than traditional ML approaches (i.e., NB, LR and RF). These results refute the original study's findings.}
\vspace{0.3cm}

\section{Threats to Validity}
\label{sec:threats}
We mitigated \textit{construct validity} threats that may arise from the choice of the data and the way it has been collected, by using of publicly available datasets that have been carefully curated and used in previous work \cite{petric2016jinx,yatish2019mining}.
In relation to \textit{conclusion validity}, we carefully calculated the performance measures and applied statistical tests, verifying all the required assumptions. We use a robust evaluation measure (i.e., MCC) to evaluate the performance of the prediction models \cite{moussa2022issta}.
The \textit{conclusion drawn from our replication} may be affected by the fact that we use different hold-out data splits as this information was not present in the original study, however to mitigate this bias we run the experiments 30 times and report the average results herein. Similarly, the tools used to run the experiments may differ given that the original study did not report this information and the authors could not provide additional details. Also, we include different benchmarks with respect to those used in the original study, as it is preferable to use basic and widely used classifiers rather than more complex variants, based on the rationale that any novel approach should be able to outperform basic ones \cite{shepperd2012evaluating,sarro2018linear}.
The \textit{external validity} of our study can be biased by the ML techniques and subjects we considered. However, we have designed our study aiming at using ML techniques and datasets, which are as representative as possible of the defect prediction literature.  We considered traditional two-class classification techniques widely used in previous studies~\cite{hall2011systematic} as our aim is to benchmark one-class predictors vs. traditional two-class predictors, and not to search for the best prediction technique. If OCSVM is not able to outperform such traditional baselines, it is reasonable to assume that it will not perform better than more sophisticated ones proposed for cross-version and cross-project defect prediction (e.g., \cite{Nam2013,Xia2016,Herbold2018,Zhou18,Hosseini2019,amasaki2020cross}).
Moreover, we used techniques freely available from a popular API library (i.e., {\tt Scikit-Learn}) to mitigate any bias/error arising from ad-hoc implementations, however different libraries may bring different results\cite{moussa2022granted}. We also used publicly available datasets previously used in the literature, which are of different nature and size, and have been carefully curated in previous work as explained in Section \ref{sec:data}. We cannot claim the results will generalise to other software, despite the fact that we have analysed the use of the proposed approaches for 15 real-world software projects having different characteristics and in three different scenarios (hold-out, cross-release, cross-project). The only way to mitigate this threat is to replicate the present study on other datasets. In order to allow and facilitate future replications and extensions of this work, we will make the code and data publicly available upon acceptance.


\section{Related Work}
\label{sec:relatedWork}
A great deal of research has been conducted to predict defect in software modules. This includes work that explores a wide range of two-class classifiers as potential solutions to identify the possibility of a module being defective. A survey of this can be found elsewhere \cite{hall2011systematic}. 
However, only few studies investigate one-class classifiers or anomaly detection approaches to predict software defects. 
The work by \originalp, replicated herein, is the first using an ML classifier (i.e., OCSVM) which learns solely from defective training data. The work by Ding et al. \cite{ding2019novel} uses Isolation Forest, which instead uses some amount of defective data, together with non-defective one, to build the prediction model. Their results, obtained 
on five NASA datasets \cite{Sayyad-Shirabad+Menzies:2005}, show that on average over all datasets, Isolation Forest obtained higher F-Measure and AUC values than Bagging, Boosting, and RF. Their results also show that the use of a certain amount of defective data, together with non-defective ones, can improve defect prediction performance, which is in line with what we observed by using more recent datasets and other validation scenarios. Lomio et al. \cite{10011488} conducted an empirical investigation on 32 open-source projects evaluating three anomaly detection methods, including OCSVM, for fine-grained just-in-time defect prediction. Their results do not reveal significant advantages that justify the benefit of using anomaly detection over more traditional machine learning approaches. These results are in line with those we obtained by using OCSVM at a different granularity level (i.e., whithin-project, cross-version, cross-project).

\section{Lessons Learnt and Future Work}
\label{sec:conclusions&futurework}
In this paper we have investigated the effectiveness of OCSVM for software defect prediction by carrying out a comprehensive empirical study involving the most commonly used machine learners and evaluation scenarios (i.e., within-project, cross-version, cross-project).
We summarise the main lessons learnt below: 

\begin{itemize}

\item Overall all classifiers investigated herein perform poorly according to MCC. Even if there are statistical significance differences among some of them, the overall effect has no practical applications. Thus, we are not able to recommend any of these classifiers to a practitioner dealing with software defect data similar to the one investigated herein.

\item OCSVM does not pass the sanity check for the within-project scenario (i.e., its estimates are significantly better than random guessing in only 17\% of the cases) and is not as effective as SVM, SVM${_T}$ and the other traditional two-class classifiers (i.e., NB, LR, RF) for this scenario.

\item OCSVM passes the sanity check for both the cross-version and cross-project scenarios (i.e., its estimates are always statistically significant better than random guessing), it also provides significantly better estimates than SVM in all cases, and than SVM$_{T}$ in 11\% of the cases for cross-version and 22\% of the cases for cross-project. On the other hand, it is able to outperform with statistical significant results the traditional two-class classifiers (i.e., NB, LR, RF) in 37\% and 48\% of the cases in total for the cross-version scenario and cross-project scenario, respectively.
\end{itemize}

\noindent When we consider OCVSM$_{T}$, which makes use of a minimal number of defective instances for hyper-parameter tuning, the overall results improve for all scenarios, as follows:

\begin{itemize}

\item 
OCVSM$_{T}$ performs statistically significantly better than random guessing in all cases for the within-project scenario (which is a very good improvement compared to the results of OCSVM for this same scenario) passing the sanity check.
Similarly, its performance against SVM, SVM$_{T}$ and the other traditional classifiers (i.e., NB, LR, RF) improves to 43\% of the cases (compared to the 20\% of OCSVM). Overall, we conclude that OCVSM$_{T}$ is effective in less than half of the cases for within-project defect prediction.

\item OCVSM$_{T}$ performs statistically significantly better than random guessing and SVM in all cases for both the cross-version and cross-project scenarios. Whereas it is significantly better than SVM$_{T}$ in 55\% of the cases for the cross-version scenario and 100\% of the cases for the cross-project scenario.
OCVSM$_{T}$ also performs statistically significantly better than the traditional ML in 56\% of the cases for cross-version and 81\% of the cases for cross-project.

\end{itemize}

The results, overall, suggest that neither OCSVM nor OCSVM$_T$ is as effective as the traditional two-class classifiers for the within-project scenario. Therefore their use cannot be recommended in this case, with the only exception being that OCSVM$_T$ should be preferred to SVM when it is not feasible to tune the latter. On the other hand, we observe that while OCSVM also remains ineffective for cross-version and cross-project defect prediction, its tuned counterpart (i.e., OCSVM$_T$) achieves statistically significantly better results than traditional approaches in 64\% of the cases for cross-version and in 67\% of the cases for cross-projects. 

Although our study reveals negative results for OCSVM (i.e., OCSVM is not able to consistently outperform the more traditional two-class classifiers and RF is overall the best performing approach according to RQ3), we believe the results also shed light on an another interesting aspect. In fact, the above findings suggest that when the training data is more heterogeneous, using OCSVM tuned with a minimal number of defective instances, can improve the estimates with respect to using traditional approaches trained on all the available defective and non-defective instances.
This points to recommendations for future work to explore the extent to which performing hyper-parameter tuning on different ratios of non-defective instances affects the prediction performance.

\section*{Data Availability}
The data and the code used in this work are available at \url{https://solar.cs.ucl.ac.uk/os.html}.

\ifCLASSOPTIONcompsoc
  \section*{Acknowledgments}
\else
  \section*{Acknowledgment}
\fi
This research is supported by the ERC Advanced fellowship grant no. 741278 and by the Lebanese American University under fund no. SRDC-F-2018-107. 

\ifCLASSOPTIONcaptionsoff
  \newpage
\fi


\bibliographystyle{IEEEtran}
\bibliography{references.bib}

\begin{thebibliography}{10}
\providecommand{\url}[1]{#1}
\csname url@samestyle\endcsname
\providecommand{\newblock}{\relax}
\providecommand{\bibinfo}[2]{#2}
\providecommand{\BIBentrySTDinterwordspacing}{\spaceskip=0pt\relax}
\providecommand{\BIBentryALTinterwordstretchfactor}{4}
\providecommand{\BIBentryALTinterwordspacing}{\spaceskip=\fontdimen2\font plus
\BIBentryALTinterwordstretchfactor\fontdimen3\font minus
  \fontdimen4\font\relax}
\providecommand{\BIBforeignlanguage}[2]{{%
\expandafter\ifx\csname l@#1\endcsname\relax
\typeout{** WARNING: IEEEtran.bst: No hyphenation pattern has been}%
\typeout{** loaded for the language `#1'. Using the pattern for}%
\typeout{** the default language instead.}%
\else
\language=\csname l@#1\endcsname
\fi
#2}}
\providecommand{\BIBdecl}{\relax}
\BIBdecl

\bibitem{Capers2008}
C.~Jones, \emph{Applied Software Measurement: Global Analysis of Productivity
  and Quality}, 3rd~ed.\hskip 1em plus 0.5em minus 0.4em\relax McGraw-Hill
  Education Group, 2008.

\bibitem{hall2011systematic}
T.~Hall, S.~Beecham, D.~Bowes, D.~Gray, and S.~Counsell, ``A systematic
  literature review on fault prediction performance in software engineering,''
  \emph{IEEE TSE}, vol.~38, no.~6, pp. 1276--1304, 2011.

\bibitem{breiman2001random}
L.~Breiman, ``Random forests,'' \emph{Machine learning}, vol.~45, no.~1, pp.
  5--32, 2001.

\bibitem{burges1998tutorial}
C.~J. Burges, ``A tutorial on support vector machines for pattern
  recognition,'' \emph{Data mining and knowledge discovery}, vol.~2, no.~2, pp.
  121--167, 1998.

\bibitem{witten2005practical}
I.~H. Witten, E.~Frank, and M.~A. Hall, ``Practical machine learning tools and
  techniques,'' \emph{Morgan Kaufmann}, p. 578, 2005.

\bibitem{yao2020assessing}
J.~Yao and M.~Shepperd, ``Assessing software defection prediction performance:
  why using the matthews correlation coefficient matters,'' in \emph{Procs. of
  the Evaluation and Assessment in Software Engineering}, 2020, pp. 120--129.

\bibitem{chawla2002smote}
N.~V. Chawla, K.~W. Bowyer, L.~O. Hall, and W.~P. Kegelmeyer, ``Smote:
  synthetic minority over-sampling technique,'' \emph{Journal of artificial
  intelligence research}, vol.~16, pp. 321--357, 2002.

\bibitem{he2008adasyn}
H.~He, Y.~Bai, E.~A. Garcia, and S.~Li, ``Adasyn: Adaptive synthetic sampling
  approach for imbalanced learning,'' in \emph{2008 IEEE international joint
  conference on neural networks (IEEE world congress on computational
  intelligence)}.\hskip 1em plus 0.5em minus 0.4em\relax IEEE, 2008, pp.
  1322--1328.

\bibitem{agrawal2018better}
A.~Agrawal and T.~Menzies, ``Is" better data" better than" better data
  miners"?'' in \emph{ICSE}.\hskip 1em plus 0.5em minus 0.4em\relax IEEE, 2018,
  pp. 1050--1061.

\bibitem{tantithamthavorn2018impact}
C.~Tantithamthavorn, A.~E. Hassan, and K.~Matsumoto, ``The impact of class
  rebalancing techniques on the performance and interpretation of defect
  prediction models,'' \emph{IEEE TSE}, vol.~46, no.~11, pp. 1200--1219, 2018.

\bibitem{manevitz2001one}
L.~M. Manevitz and M.~Yousef, ``One-class svms for document classification,''
  \emph{Journal of machine Learning research}, vol.~2, no. Dec, pp. 139--154,
  2001.

\bibitem{erfani2016high}
S.~M. Erfani, S.~Rajasegarar, S.~Karunasekera, and C.~Leckie,
  ``High-dimensional and large-scale anomaly detection using a linear one-class
  svm with deep learning,'' \emph{Pattern Recognition}, vol.~58, pp. 121--134,
  2016.

\bibitem{zhang2006fall}
T.~Zhang, J.~Wang, L.~Xu, and P.~Liu, ``Fall detection by wearable sensor and
  one-class svm algorithm,'' in \emph{Intelligent computing in signal
  processing and pattern recognition}.\hskip 1em plus 0.5em minus 0.4em\relax
  Springer, 2006, pp. 858--863.

\bibitem{li2010positive}
W.~Li, Q.~Guo, and C.~Elkan, ``A positive and unlabeled learning algorithm for
  one-class classification of remote-sensing data,'' \emph{IEEE Transactions on
  Geoscience and Remote Sensing}, vol.~49, no.~2, pp. 717--725, 2010.

\bibitem{khan2014one}
S.~S. Khan and M.~G. Madden, ``One-class classification: taxonomy of study and
  review of techniques,'' \emph{The Knowledge Engineering Review}, vol.~29,
  no.~3, pp. 345--374, 2014.

\bibitem{chen2016software}
L.~Chen, B.~Fang, and Z.~Shang, ``Software fault prediction based on one-class
  svm,'' in \emph{ICMLC}, vol.~2.\hskip 1em plus 0.5em minus 0.4em\relax IEEE,
  2016, pp. 1003--1008.

\bibitem{10011488}
F.~Lomio, L.~Pascarella, F.~Palomba, and V.~Lenarduzzi, ``Regularity or
  anomaly? on the use of anomaly detection for fine-grained jit defect
  prediction,'' in \emph{2022 48th Euromicro Conference on Software Engineering
  and Advanced Applications (SEAA)}, 2022, pp. 270--273.

\bibitem{shull2008role}
F.~J. Shull, J.~C. Carver, S.~Vegas, and N.~Juristo, ``The role of replications
  in empirical software engineering,'' \emph{Empirical software engineering},
  vol.~13, no.~2, pp. 211--218, 2008.

\bibitem{paige2017foreword}
R.~F. Paige, J.~Cabot, and N.~A. Ernst, ``Foreword to the special section on
  negative results in software engineering,'' 2017.

\bibitem{tichy2000hints}
W.~F. Tichy, ``Hints for reviewing empirical work in software engineering,''
  \emph{Empirical Software Engineering}, vol.~5, no.~4, pp. 309--312, 2000.

\bibitem{ferguson2012vast}
C.~J. Ferguson and M.~Heene, ``A vast graveyard of undead theories: Publication
  bias and psychological science’s aversion to the null,'' \emph{Perspectives
  on Psychological Science}, vol.~7, no.~6, pp. 555--561, 2012.

\bibitem{menzies2017negative}
T.~Menzies, Y.~Yang, G.~Mathew, B.~Boehm, and J.~Hihn, ``Negative results for
  software effort estimation,'' \emph{Empirical Software Engineering}, vol.~22,
  no.~5, pp. 2658--2683, 2017.

\bibitem{petric2016jinx}
J.~Petri{\'c}, D.~Bowes, T.~Hall, B.~Christianson, and N.~Baddoo, ``The jinx on
  the nasa software defect data sets,'' in \emph{Int. Conference on Evaluation
  and Assessment in Software Engineering}, 2016, pp. 1--5.

\bibitem{yatish2019mining}
S.~Yatish, J.~Jiarpakdee, P.~Thongtanunam, and C.~Tantithamthavorn, ``Mining
  software defects: should we consider affected releases?'' in
  \emph{ICSE}.\hskip 1em plus 0.5em minus 0.4em\relax IEEE, 2019, pp. 654--665.

\bibitem{d2012evaluating}
M.~D’Ambros, M.~Lanza, and R.~Robbes, ``Evaluating defect prediction
  approaches: a benchmark and an extensive comparison,'' \emph{EMSE}, vol.~17,
  no. 4-5, pp. 531--577, 2012.

\bibitem{sarro2018linear}
F.~Sarro and A.~Petrozziello, ``Linear programming as a baseline for software
  effort estimation,'' \emph{ACM TOSEM}, vol.~27, no.~3, pp. 1--28, 2018.

\bibitem{scikitLearn}
F.~Pedregosa, G.~Varoquaux, A.~Gramfort, V.~Michel, B.~Thirion, O.~Grisel,
  M.~Blondel, P.~Prettenhofer, R.~Weiss, V.~Dubourg, J.~Vanderplas, A.~Passos,
  D.~Cournapeau, M.~Brucher, M.~Perrot, and E.~Duchesnay, ``Scikit-learn:
  Machine learning in {P}ython,'' \emph{Journal of Machine Learning Research},
  vol.~12, pp. 2825--2830, 2011.

\bibitem{sarro2012further}
F.~Sarro, S.~Di~Martino, F.~Ferrucci, and C.~Gravino, ``A further analysis on
  the use of genetic algorithm to configure support vector machines for
  inter-release fault prediction,'' in \emph{Procs. of ACM SAC}, 2012, pp.
  1215--1220.

\bibitem{di2011genetic}
S.~Di~Martino, F.~Ferrucci, C.~Gravino, and F.~Sarro, ``A genetic algorithm to
  configure support vector machines for predicting fault-prone components,'' in
  \emph{Int. conference on product focused software process improvement}.\hskip
  1em plus 0.5em minus 0.4em\relax Springer, 2011, pp. 247--261.

\bibitem{shepperd2013data}
M.~Shepperd, Q.~Song, Z.~Sun, and C.~Mair, ``Data quality: Some comments on the
  nasa software defect datasets,'' \emph{IEEE TSE}, vol.~39, no.~9, pp.
  1208--1215, 2013.

\bibitem{herbold2017comparative}
S.~Herbold, A.~Trautsch, and J.~Grabowski, ``A comparative study to benchmark
  cross-project defect prediction approaches,'' \emph{IEEE TSE}, vol.~44,
  no.~9, pp. 811--833, 2017.

\bibitem{nam2017heterogeneous}
J.~Nam, W.~Fu, S.~Kim, T.~Menzies, and L.~Tan, ``Heterogeneous defect
  prediction,'' \emph{IEEE TSE}, vol.~44, no.~9, pp. 874--896, 2017.

\bibitem{moussa2022issta}
R.~Moussa and F.~Sarro, ``On the use of evaluation measures for defect
  prediction studies,'' in \emph{31st ACM SIGSOFT Int. Symposium on Software
  Testing and Analysis (ISSTA 2022)}.\hskip 1em plus 0.5em minus 0.4em\relax
  ACM, 2022.

\bibitem{shepperd2014researcher}
M.~Shepperd, D.~Bowes, and T.~Hall, ``Researcher bias: The use of machine
  learning in software defect prediction,'' \emph{IEEE TSE}, vol.~40, no.~6,
  pp. 603--616, 2014.

\bibitem{woolson2007wilcoxon}
R.~Woolson, ``Wilcoxon signed-rank test,'' \emph{Wiley encyclopedia of clinical
  trials}, pp. 1--3, 2007.

\bibitem{sarro2016multi}
F.~Sarro, A.~Petrozziello, and M.~Harman, ``Multi-objective software effort
  estimation,'' in \emph{Procs. of ICSE}.\hskip 1em plus 0.5em minus
  0.4em\relax IEEE, 2016, pp. 619--630.

\bibitem{ArcuriB14}
A.~Arcuri and L.~Briand, ``A hitchhiker's guide to statistical tests for
  assessing randomized algorithms in software engineering,'' \emph{STVR},
  vol.~24, no.~3, pp. 219--250, 2014.

\bibitem{arcuri2014hitchhiker}
------, ``A hitchhiker's guide to statistical tests for assessing randomized
  algorithms in software engineering,'' \emph{Software Testing, Verification
  and Reliability}, vol.~24, no.~3, pp. 219--250, 2014.

\bibitem{harman2014less}
M.~Harman, S.~Islam, Y.~Jia, L.~L. Minku, F.~Sarro, and K.~Srivisut, ``Less is
  more: Temporal fault predictive performance over multiple hadoop releases,''
  in \emph{Int. Symposium on Search Based Software Engineering}.\hskip 1em plus
  0.5em minus 0.4em\relax Springer, 2014, pp. 240--246.

\bibitem{ahluwalia2020need}
A.~Ahluwalia, M.~Di~Penta, and D.~Falessi, ``On the need of removing last
  releases of data when using or validating defect prediction models,''
  \emph{arXiv preprint arXiv:2003.14376}, 2020.

\bibitem{jimenez2019importance}
M.~Jimenez, R.~Rwemalika, M.~Papadakis, F.~Sarro, Y.~Le~Traon, and M.~Harman,
  ``The importance of accounting for real-world labelling when predicting
  software vulnerabilities,'' in \emph{Procs. of ESEC/FSE}, 2019, pp. 695--705.

\bibitem{bangash2020time}
A.~A. Bangash, H.~Sahar, A.~Hindle, and K.~Ali, ``On the time-based conclusion
  stability of cross-project defect prediction models,'' \emph{Empirical
  Software Engineering}, vol.~25, no.~6, pp. 5047--5083, 2020.

\bibitem{lavalley2008logistic}
M.~P. LaValley, ``Logistic regression,'' \emph{Circulation}, vol. 117, no.~18,
  pp. 2395--2399, 2008.

\bibitem{tantithamthavorn2016automated}
C.~Tantithamthavorn, S.~McIntosh, A.~E. Hassan, and K.~Matsumoto, ``Automated
  parameter optimization of classification techniques for defect prediction
  models,'' in \emph{ICSE}, 2016, pp. 321--332.

\bibitem{Sayyad-Shirabad+Menzies:2005}
\BIBentryALTinterwordspacing
J.~Sayyad~Shirabad and T.~Menzies, ``{The {PROMISE} Repository of Software
  Engineering Databases.}'' School of Information Technology and Engineering,
  University of Ottawa, Canada, 2005. [Online]. Available:
  \url{http://promise.site.uottawa.ca/SERepository}
\BIBentrySTDinterwordspacing

\bibitem{shepperd2012evaluating}
M.~Shepperd and S.~MacDonell, ``Evaluating prediction systems in software
  project estimation,'' \emph{Information and Software Technology}, vol.~54,
  no.~8, pp. 820--827, 2012.

\bibitem{Nam2013}
J.~Nam, S.~J. Pan, and S.~Kim, ``Transfer defect learning,'' in \emph{ICSE},
  2013, pp. 382--391.

\bibitem{Xia2016}
X.~Xia, D.~Lo, S.~J. Pan, N.~Nagappan, and X.~Wang, ``Hydra: Massively
  compositional model for cross-project defect prediction,'' \emph{IEEE TSE},
  vol.~42, no.~10, pp. 977--998, 2016.

\bibitem{Herbold2018}
S.~Herbold, A.~Trautsch, and J.~Grabowski, ``A comparative study to benchmark
  cross-project defect prediction approaches,'' \emph{IEEE TSE}, vol.~44,
  no.~09, pp. 811--833, sep 2018.

\bibitem{Zhou18}
\BIBentryALTinterwordspacing
Y.~Zhou, Y.~Yang, H.~Lu, L.~Chen, Y.~Li, Y.~Zhao, J.~Qian, and B.~Xu, ``How far
  we have progressed in the journey? an examination of cross-project defect
  prediction,'' \emph{ACM Trans. Softw. Eng. Methodol.}, vol.~27, no.~1, Apr.
  2018. [Online]. Available: \url{https://doi.org/10.1145/3183339}
\BIBentrySTDinterwordspacing

\bibitem{Hosseini2019}
S.~Hosseini, B.~Turhan, and D.~Gunarathna, ``A systematic literature review and
  meta-analysis on cross project defect prediction,'' \emph{IEEE TSE}, vol.~45,
  no.~2, pp. 111--147, 2019.

\bibitem{amasaki2020cross}
S.~Amasaki, ``Cross-version defect prediction: use historical data,
  cross-project data, or both?'' \emph{Empirical Software Engineering},
  vol.~25, no.~2, pp. 1573--1595, 2020.

\bibitem{moussa2022granted}
R.~Moussa and F.~Sarro, ``Do not take it for granted: Comparing open-source
  libraries for software development effort estimation,'' 2022.

\bibitem{ding2019novel}
Z.~Ding, Y.~Mo, and Z.~Pan, ``A novel software defect prediction method based
  on isolation forest,'' in \emph{2019 Int. Conference on Quality, Reliability,
  Risk, Maintenance, and Safety Engineering (QR2MSE)}.\hskip 1em plus 0.5em
  minus 0.4em\relax IEEE, 2019, pp. 882--887.

\end{thebibliography}

\end{document}